\documentclass[prb,twocolumn,showpacs,preprintnumbers]{revtex4}
\usepackage{graphicx}
\usepackage{amsmath}
\usepackage{longtable}
\begin{document}
\title{Exchange effects on electron scattering
through a quantum dot embedded in a two-dimensional semiconductor
structure}
\author{L. K. Castelano and G.-Q. Hai}\email{hai@ifsc.usp.br}
\affiliation{Instituto de F\'isica de S\~ao
Carlos, Universidade de S\~ao Paulo, 13560-970, S\~ao Carlos, SP,
Brazil}
\author{M.-T. Lee}
\affiliation{Departamento de Qu\'imica, Universidade Federal de
S\~ao Carlos, 13565-905, S\~ao Carlos, SP, Brazil}
\begin{abstract}
We have developed a theoretical method to study scattering processes
of an incident electron through an N-electron quantum dot (QD)
embedded in a two-dimensional (2D) semiconductor. The generalized
Lippmann-Schwinger equations including the electron-electron
interaction in this system are solved for the continuum electron by
using the method of continued fractions (MCF) combined with 2D
partial wave expansion technique. The method is applied to a
one-electron QD case. Scattering cross-sections are obtained for
both the singlet and triplet couplings between the incident electron
and the QD electron during the scattering. The total elastic
scattering cross-sections as well as the spin-flip scattering
cross-sections resulting from the exchange potential are presented.
Furthermore, inelastic scattering processes are also studied using a
multichannel formalism of the MCF.
\end{abstract}

\maketitle


\section{Introduction}
Electron scattering and transport through quantum dots (QDs) in a
semiconductor
nanostructure\cite{koppens,qu,fransson,konig,zhang,engel} have been
intensively studied recently. The spin-dependent transport
properties are of particular interest for its possible applications,
{\it e.g.}, the QD spin valves\cite{konig}, the quantum logic gates
using coupled QDs, as well as the spin-dependent transport in
single-electron devices\cite{seneor}, etc.. In such systems, the
electron-electron exchange potential and the electron spin states
have been utilized and
manipulated\cite{wolf,burkard,dassarma,gundogdu}. A thorough
quantitative understanding of spin-dependent properties due to
electron-electron interaction is therefore important for a
successful construction of these devices. This subject has been
investigated in different issues, such as the utilization of the
electron-electron scattering in determining the electron
entanglement dynamics\cite{busceni}, the study of spin-flip
scattering in double QDs\cite{taoji} and the scattering through a
region of nonuniform spin-orbit coupling which can form a spin
polarized beam\cite{palyi}. Theoretically the transport through QDs
has been studied by different approaches such as transfer matrix,
nonequilibrium Green's functions, random matrix theory, as well as
those methods built on the Lippmann-Schwinger (L-S) equation.

In this work, we develop a theoretical method to study electron
scattering through a quantum dot (QD) of N-electrons embedded in a
two-dimensional (2D) semiconductor system. We construct the
scattering equations including electron-electron interaction to
represent the process of a 2D free electron scattered by the QD. The
generalized multichannel Lippmann-Schwinger
equations\cite{qct,bransden} are solved for this system by using the
method of continued fractions (MCF). The MCF is an iterative method
to solve the integro-differential L-S equations, initially developed
for three-dimensional electron-atom (molecule) scattering in atomic
physics \cite{mcf}. We show that this method, combining with the
partial wave expansion technique, is of a rapid convergency for the
present problem in a 2D semiconductor system and therefore is
efficient to obtain the scattering cross-sections. As an example, we
apply this method to a one-electron QD case and obtain scattering
cross-sections resulting from both the singlet- and triplet-coupled
continuum states of two electrons (incident and QD electrons) during
the collision. The results show that the scattering processes can be
very different for singlet and triplet spin states which are mainly
originated from the different exchange interactions. From the
difference of the scattering amplitudes resulting from the singlet
and triplet couplings, we determine the spin-flip scattering
cross-sections which exhibit a maximum as a function of scattering
angle and the incident electron energy. In a multichannel
scattering, we study the inelastic scattering process in which the
incident electron is scattered by a lower energy state of the QD and
leaves behind the QD in an excited state. As expected, such an
inelastic scattering cross-section is found much smaller than the
elastic one.

This paper is organized as follows. In Sec. II we present the
Hamiltonian of the system. In Sec. III, we describe our general
theoretical approach and the one-electron QD case is given as an
example. In Sec. IV we show our numerical results for the scattering
through a one-electron QD within both the one-channel and the
multichannel models. The conclusion is presented in Sec. V. Moreover, the method of continued
fractions is briefly described in Appendix A. The 2D partial wave
expansions used in the numerical solution of the L-S equations are
presented in Appendix B.

\section{Hamiltonian of the system}\label{sed}

The system under investigation consists of an incident 2D free
electron and a quantum dot of N electrons embedded in a 2D system.
The incident electron is scattered by both the QD potential and by the
confined electrons inside the QD. The Schr\"odinger equation of the
system is given by
\begin{equation}\label{estot}
(H-\mathcal{E}_i)\Psi_i(\mathbf{\tau};\mathbf{r}_{N+1},
\sigma_{N+1})=0\; ,
\end{equation}
where $\mathbf{\tau}$ represents collectively the spatial and spin
coordinates of the N electrons localized in the QD and
$\mathbf{r}_{N+1}=(x_{N+1},y_{N+1})$ and $\sigma_{N+1}$ denote the
spatial and spin coordinates of the incident electron. The total
energy of the system is $\mathcal{E}_i$, where the subscript $i$
represents a set of quantum numbers required to specify uniquely the
initial quantum state of the system. Explicitly, the total
Hamiltonian of the system can be written as
\begin{equation}\label{estot1}
H=H_0(\mathbf{r}_{N+1})+H_{\rm QD}(\tau)+V_{\rm
int}(\mathbf{r}_1,\mathbf{r}_2,...,
    \mathbf{r}_N,\mathbf{r}_{N+1})\; ,
\end{equation}
where $H_{0}(\mathbf{r}_{N+1})=-\hbar^2\nabla^2_{N+1}/2m^*+V_{\rm
QD}(\mathbf{r}_{N+1})$, $H_{\rm QD}(\tau)$ is the Hamiltonian of the
QD of N electrons, and $V_{\rm int}$ is the interaction potential
between the incident electron at $\mathbf{r}_{N+1}$ and the N
electrons in the QD
\begin{equation}\label{vint} V_{\rm int}(\mathbf{r}_1,\mathbf{r}_2,...,
\mathbf{r}_N,\mathbf{r}_{N+1})=\frac{e^2}{\epsilon^*_0}\sum_{i=1}^{N}\frac{1}
{|\mathbf{r}_{N+1}-\mathbf{r}_i|},
\end{equation}
where $\epsilon^*_0$ is the dielectric constant of the semiconductor
material and $m^*$ is the electron effective mass. The Hamiltonian
for an unperturbed QD is given by
\begin{equation}
H_{\rm
QD}(\tau)=\sum_{i=1}^N\left(-\frac{\hbar^2}{2m^*}\nabla^2_i+V_{\rm
QD}(\mathbf{r_i})\right) +\frac{e^2}{\epsilon^*_0}\sum_{i\neq
j}^N\frac{1}{|\mathbf{r}_i-\mathbf{r}_j|} , \label{hqd}
\end{equation}
where the first term in the $rhs$ of Eq.~(\ref{hqd}) describes N
independent electrons in the QD of confinement potential $V_{\rm
QD}(\mathbf{r})$ and the second term gives the Coulomb interactions
among these electrons. The eigenenergy and eigenfunction of this
N-electron QD are denoted by $\varepsilon_n$ and $\Phi^n$,
respectively. They are determined by the following Schr\"odinger
equation
\begin{equation}
H_{\rm QD}(\tau) \Phi^n= \varepsilon_n\Phi^n , \label{ehqd}
\end{equation}
with $n=0,1,2,3$... . The ground state of the N-electron QD is
labeled by $n=0$ and the excited states by $n \ge 1$. The
eigenstates of the QD can be obtained using, {\it e.g.}, the
restricted or unrestricted Hartree-Fock (HF) methods \cite{szabo}.

\section{Scattering equations including electron exchange interaction}

In order to extract scattering properties of the system (QD +
incident electron), we can write the total wave-function $\Psi_i$ of
the system as a superposition of the QD wave-function $\Phi^n$ and
the incident electron wave-function,
\begin{equation}\label{psi}
|\Psi_i\rangle=\sum_{n=0}^{\infty}|\mathcal{A}(\Phi^n\psi_{ni})\rangle
,
\end{equation}
where $\psi_{ni}$ describes the wave-functions of the incident
(scattered) electron in the continuum states corresponding to a
quantum transition from an initial state $i$ to a final state $n$.
The operator $\mathcal{A}$ warrants the antisymmetrization property
between the QD electrons and the incident electron, defined by,
\begin{equation}
\mathcal{A}=\frac{1}{\sqrt{N+1}}\sum_{p=1}^{N+1} (-1)^{N+1-p}
\mathcal{P}_{N+1,p}
\end{equation}
where $\mathcal{P}_{N+1,p}$ is the permutation operator which
exchanges the electrons at ${\bf r}_{N+1}$ and ${\bf r}_p$. From
Eqs.~(\ref{estot}), (\ref{estot1}) and (\ref{psi}), we obtain
\begin{eqnarray}\label{es1}
&&\sum_{n=0}^{\infty}\left(-\frac{\hbar^2}{2m^*}\nabla^2_{{N+1}}+V_{\rm
QD}+H_{\rm QD}+V_{\rm int}
\right)|\mathcal{A}(\Phi^n\psi_{ni})\rangle\nonumber\\
&&=\mathcal{E}_i\sum_{n=0}^{\infty}|\mathcal{A}(\Phi^n\psi_{ni})\rangle .
\end{eqnarray}
The total energy of the system (the incident electron + QD)
$\mathcal{E}_i$ is composed by two parts. The first part is the
kinetic energy of the incident (scattering) electron and the second
is the energy of the N-electron QD in a particular configuration,
{\it i.e.}, $\mathcal{E}_i=\frac{\hbar^2k_i^2}{2m^*}+\varepsilon_i=
\frac{\hbar^2k_n^2}{2m^*}+\varepsilon_n$, for different eigenstates
of the QD ($i,n=0,1,2,...$) or different scattering channels. These
different channels appear because the incident electron can probably
be scattered inelastically, leaving the QD in a different state from
its initial. A projection of Eq.~(\ref{es1}) onto a particular QD
state $|\Phi^m\rangle$ leads to the following scattering equation
for the incident electron,
\begin{equation}\label{es4}
\frac{\hbar^2}{2m^*}\left(\nabla^2+k_m^2
\right)\psi_{mi}(\mathbf{r})=\sum_{n=0}^{\infty}V_{mn}(\mathbf{r})\psi_{ni}(\mathbf{r})
\end{equation}
for $i,m=0,1,2,...$, where $\mathbf{r}=\mathbf{r}_{N+1}$ and
$V_{mn}=V_{mn}^{\rm st}+ V_{mn}^{\rm ex}$ with $V_{mn}^{\rm st}$ the
static potential and $V_{mn}^{\rm ex}$ the exchange potential due
the nonlocal interaction, giving by
\begin{equation}
V_{mn}^{\rm st}(\mathbf{r})= V_{\rm QD}(\mathbf{r})\delta_{mn} +
\frac{e^2}{\epsilon_0^*}\sum_{j=1}^N
\langle\Phi^{m}|\frac{e^{-\lambda|\mathbf{r}-\mathbf{r_j}|}}
    {|\mathbf{r}-\mathbf{r_j}|}|
    \Phi^{n}\rangle ,\label{vest}
\end{equation}
and
\begin{eqnarray}
V_{mn}^{\rm ex}(\mathbf{r})\psi_{ni}(\mathbf{r}) &=&
(H_{0}(\mathbf{r})-\frac{\hbar^2k_m^2}{2m^*}) \langle\Phi^{m}
|\mathcal{A}'(\Phi^n\psi_{ni})\rangle\nonumber \\
&+&\frac{e^2}{\epsilon_0^*}\sum_{j=1}^N \langle\Phi^{m}|\frac{1}
    {|\mathbf{r}-\mathbf{r_j}|}|\mathcal{A}'(\Phi^n\psi_{ni})\rangle , \label{vex}
\end{eqnarray}
respectively, where
$\mathcal{A}'=\sum_{p=1}^{N}(-1)^{N+1-p}\mathcal{P}_{N+1,p}$. In
Eq.~(\ref{vest}) we have introduced a screening $e^{-\lambda
|\bf{r}-\bf{r'}|}$ on the direct Coulomb potential for two reasons:
(i) the the ionized impurities in the semiconductor nanostructure
and/or the external electrodes screen the direct Coulomb potential
and (ii) at the $|\mathbf{r}|\rightarrow\infty$ limit the scattering
potential should decay faster than $1/|\mathbf{r}|$. The screening
length is given by $\lambda^{-1}$. Notice that we do not consider
the screening on the exchange potential because this potential is
non-zero inside the QD only. Inclusion of the screening on the
exchange potential in Eq.~(\ref{vex}) is possible but it will not
affect much our results and complicates the numerical calculation.

The scattering equation is a system of coupled integro-differential
equations. The corresponding generalized L-S equation for such a
multichannel scattering problem is given by
\begin{eqnarray}
&& \psi_{mi}(\mathbf{r})=\varphi_{i}(\mathbf{r})\delta_{mi} \nonumber\\
&& +\sum_{n=0}^{\infty}\int d{\bf r}'
G^{(0)}(\mathbf{k}_m,\mathbf{r},\mathbf{r}')V_{mn}(\mathbf{r}')
\psi_{ni}(\mathbf{r}') ,\label{elsmc}\\
&& \;\;\mathrm{for}\; \; i, m=0,1,2\dots \nonumber
\end{eqnarray}
with an incident plane wave
$\varphi_{i}(\mathbf{r})=e^{i\mathbf{k}_i.\mathbf{r}}=e^{ik_i x}$ in
the $x$-direction. The Green's function
$G^{(0)}(\mathbf{k},\mathbf{r},\mathbf{r}')$ in the above equation
is
\begin{equation}
G^{(0)}(\mathbf{k},\mathbf{r},\mathbf{r'})
=-\frac{2m^*}{\hbar^2}(i/4)H_0^{(1)}(k|\mathbf{r}-\mathbf{r'}|),
\label{fgreen}
\end{equation}
where $H_0^{(1)}$ is the usual zero order Hankel's
function\cite{morse}.

At $|\mathbf{r}|\rightarrow\infty$ limit, the asymptotic form of
Eq.~(\ref{elsmc}) for the scattered wave-function in a 2D system is
given by
\begin{equation}\label{elsmcass}
\psi_{mi}(\mathbf{r})\hbox{\space \raise-2mm\hbox{$\textstyle
    \longrightarrow \atop \scriptstyle |\mathbf{r}|\rightarrow\infty$} \space}
    e^{ik_i x}\delta_{mi}+\frac{2m^*}{\hbar^2}\sqrt{\frac{i}{k_m}}\frac{e^{+
ik_mr }}{\sqrt{r}}f_{k_m,k_i}(\theta) ,
\end{equation}
where $f_{k_m,k_i}(\theta)$ is the scattering amplitude
\begin{equation}\label{amp}
  f_{k_m,k_i}(\theta)=-\frac{1}{4}\sqrt{\frac{2}{\pi}}\langle \mathbf{k}_m|T(E)
  |\mathbf{k}_i\rangle
\end{equation}
with
\begin{equation}
\langle \mathbf{k}_m|T(E)|\mathbf{k}_i \rangle=\sum_{n=0}^{\infty}\int d{\bf r}'
e^{-
i\mathbf{k}_m.\mathbf{r'}}V_{mn}(\mathbf{r'})\psi_{ni}(\mathbf{r'}).\nonumber
\end{equation}
The momenta of the initial and final states of the incident
(scattered) electron are $\mathbf{k}_i$ and $\mathbf{k}_m$,
respectively, and $\theta$ is the scattering angle between them. It
is evident from Eq.~(\ref{elsmc}) and its boundary condition
Eq.~(\ref{elsmcass}) that the different scattering channels are
coupled to each other through the interaction potential $V_{mn}$.

In the above procedure in dealing with the electron scattering
through a QD, both the electron-electron exchange and correlation
interactions are presented. However, it is difficult to include
a complete correlation effect in a practical calculation. For that,
 besides an exact solution for the N-electron QD, a full sum over
all the intermediate states $n$ in the scattering equation
[Eq.~(\ref{es4})] is needed, which is a formidable task in a
self-consistent calculation. In an alternative way, the correlation
effects can be considered by adding an effective correlation
potential in the scattering equation\cite{qct}. In the present work,
we focus on the exchange effects in the scattering process and limit
the sum over $n$ to a few lowest energy levels of the QD. For this
reason, we prefer to call the nonlocal interaction potential
$V_{mn}^{\rm ex}$ in Eq.~(\ref{vex}) as exchange potential though
the correlation are partially included in a multichannel treatment.

The differential cross-section (DCS) for a scattering from initial
state $i$ ({\it i.e.} the incident electron of kinetic energy
$E_i=\frac{\hbar^2k_i^2}{2m^*}$ and the QD in the state
$\varepsilon_i$) to final state $m$  ({\it i.e.}
$E_m=\frac{\hbar^2k_m^2}{2m^*}$ and the QD in the state $m$) is
given by
\begin{equation}\label{scdmc}
\sigma_{mi} (\theta)=\frac{k_m}{k_i^2}|f_{k_m,k_i}(\theta)|^2 .
\end{equation}
The integral cross-section (ICS) which is an energy dependent
quantity can be found by
\begin{equation}\label{scimc}
\Gamma_{mi} (E_i)=\int_0^{2\pi}\sigma_{mi} (\theta)d\theta .
\end{equation}
When the state of the QD remains the same $i.e.$ ($m=i$), before and after the scattering,
  the process is
called elastic. Otherwise, the scattering is inelastic. A possible
scattering is the so-called super-elastic scattering ($E_m>E_i$)
where the incident electron is scattered out with a higher energy by
an QD initially in an excited state. Because the different
scattering channels are coupled to each other, we have to solve the
multichannel L-S equation to obtain the scattering probabilities
through different channels simultaneously for the same total energy
of the system.

There are different numerical methods to solve the above coupled
integro-differential L-S equations. In this work, we use the
so-called method of continued fractions (MCF, see Appendix A) which
was originally developed in three-dimensional formulation for
electron-atom \cite{mcf,lee} and electron-molecule \cite{lee1}
scatterings at the single- and multi-channel level of
approximations. Here, we apply this method to electron-QD scattering
in a two-dimensional semiconductor system. The MCF is an iterative
method to solve the L-S equation. The advantage of this method lies
on its rapid convergency and its unnecessity of a basis function for
expansion of the continuum wave-functions. Using the MCF, we can
obtain the $\mathbf{T}$ matrix and consequently the DCS according to
Eqs.~(\ref{amp}) and (\ref{scdmc}). The two-dimensional integrations
on the interaction potentials in Eqs.~(\ref{vest}), (\ref{vex}), and
(\ref{elsmc}) are simplified by using partial wave expansion which
is described in Appendix B.

\subsection{One-channel approximation}

When a quantum dot is initially in its ground state and keeps in the
same state after the collision, the scattering is elastic and the
scattering process associated to the ground state of the QD is of
dominant contribution to the scattering cross-section. In this case,
one-channel treatment can be a reasonably good approximation to
calculate scattering cross-section even if the incident electron is
of enough energy and thus several inelastic channels are open during
the collision. When only the elastic channel is considered ({\it
i.e.} $i=m=n=0$), Eq.~(\ref{es4}) is reduced to
\begin{equation}\label{es5}
\frac{\hbar^2}{2m^*}\left(\nabla^2_{\mathbf{r}}+k^2
\right)\psi(\mathbf{r})=V(\mathbf{r})\psi(\mathbf{r}),
\end{equation}
where $\psi(\mathbf{r})=\psi_{00}(\mathbf{r})$,
$V(\mathbf{r})=V_{00}(\mathbf{r})$ and $k=k_0$. The L-S equation for
the scattered electron becomes
\begin{equation}\label{els}
\psi(\mathbf{r})=\varphi_\mathbf{k}(\mathbf{r})+\int d{\bf r}'
G^{(0)}(\mathbf{k},\mathbf{r},\mathbf{r}')V(\mathbf{r}')\psi(\mathbf{r}')
,
\end{equation}
where $\varphi_{\mathbf{k}}(\mathbf{r})=e^{ikx}$ and the Green's
function is given by Eq.~(\ref{fgreen}).

When the scattering potential in the above equation is central, {\it
i.e.}, $V(\mathbf{r})=V(r)$, the L-S equation can be solved easily
using the partial wave expansion technique as described in Appendix
B. Moreover, the scattering amplitude or the cross-section in this
case can be obtained in terms of the phase-shifts of the partial
waves. The DCS as a function of partial wave phase-shift $\Delta_l$
is given as:
\begin{eqnarray}
&&\sigma_{00} (\theta)=\frac{1}{k}\left|f_{k,k}(\theta)\right|^2
\nonumber \\
&& = {2\over \pi k} \left| \sum_{l=0}^\infty
\kappa_le^{i\Delta_l}\sin\Delta_l \cos(l\theta)\right| ^2,
\nonumber
\end{eqnarray}
and the ICS is given by
\begin{eqnarray}
\Gamma_{00}(E_0)=\frac{4}{k}\sum_{l=0}^\infty
\kappa_l\sin^2\Delta_l\; ,\nonumber \label{scps}
\end{eqnarray}
where $\kappa_l=1$ for $l=0$ and $\kappa_l=2$ for $l\neq0$.

\subsection{Scattering by a one-electron QD}

Electron scattering and transport through a QD of a few electrons
are currently of great experimental and theoretical interest. Here
we present the case of a QD with only one confined electron. We
focus on the exchange effect on the electron scattering and the
spin-flip scattering mechanism. The total Hamiltonian
Eq.~(\ref{estot1}) in the one-electron QD case is given by,
\begin{equation}
H(\mathbf{r}_1,\mathbf{r}_2)=\frac{-\hbar^2}{2m^*}\nabla^2_2+ V_{\rm
QD}(\mathbf{r}_{2}) +H_{\rm QD}(\mathbf{r}_{1})+V_{\rm
int}(\mathbf{r}_1,\mathbf{r}_2)\label{htot},
\end{equation}
with
\begin{equation}
V_{\rm int}(\mathbf{r}_1,\mathbf{r}_2)
=\frac{e^2}{\epsilon_0^*}\frac{1}
    {|\mathbf{r}_{2}-\mathbf{r}_1|},
\end{equation}
where $\mathbf{r}_1$ labels the localized electron in the QD and
$\mathbf{r}_2$ the incident electron. As we have mentioned, to solve
the scattering problem, we need firstly to know the electron states
in the QD which are determined by the following equation,
\begin{equation}
H_{\rm QD}(\mathbf{r})\zeta^n(\mathbf{r})=
\left[-\frac{\hbar^2}{2m^*}\nabla^2+V_{\rm QD}(\mathbf{r})
\right]\zeta^n(\mathbf{r})=\varepsilon_n\zeta^n(\mathbf{r}).
\label{hqdN1}
\end{equation}
The solution of this one-electron QD is straightforward as soon as
the confinement potential $V_{\rm QD}$ is defined.

According to Eq.~(\ref{es4}), there is an infinite number of quantum
states involved in the scattering. In performing a numerical
calculation, however, we have to truncate this to a finite number of
states. As a matter of fact, when the QD is initially in its ground
state and the incident electron has a small kinetic energy, it is a
good approximation to consider only a few scattering channels
associated to the low-energy levels of the QD. In the present
calculation, we consider the channels associated to the ground state
$\varepsilon_0$ and two excited states $\varepsilon_1$ and
$\varepsilon_2$ of the QD. When the incident electron passes through
the QD initially in the ground state $\varepsilon_0$, the scattering
can be either elastic keeping the QD in the same state or inelastic
leaving behind the QD in an excited state. According to
Eq.~(\ref{psi}), the spatial part of the total wave-function of the
two electrons (one incident and the other confined) can be written
as, within a three-level model,
\begin{equation}
\Psi_i(\mathbf{r}_1,\mathbf{r}_2)=\sum_{n=0}^2
\left[\zeta^n(\mathbf{r}_1) \psi_{ni}(\mathbf{r}_2)\pm
\zeta^n(\mathbf{r}_2) \psi_{ni}(\mathbf{r}_1)\right], \label{wfsc}
\end{equation}
where the signs ($+$) and ($-$) represent the singlet and triplet
total-spin states of the two-electron system, respectively. The
scattering equation [Eq.~(\ref{es4})] becomes,
\begin{equation}
\frac{\hbar^2}{2m^*}\left(\nabla^2_{2}+k^2_m
\right)\psi_{mi}(\mathbf{r_2})= \sum_{n=0}^2[V_{mn}^{\rm
st}(\mathbf{r}_{2}) \pm V_ {mn}^{\rm ex}(\mathbf{r}_{2})]
\psi_{ni}(\mathbf{r_2})\label{eses}
\end{equation}
with
\begin{equation}
V^{\rm st}_{mn}=V_{\rm QD}\delta_{mn}+\langle\zeta^m|V_{\rm
int}|\zeta^n\rangle
\end{equation}
and
\begin{equation}
V^{\rm ex}_{mn}\psi_{ni}= \langle\zeta^m|V_{\rm
int}|\psi_{ni}\rangle\zeta^n
+(\varepsilon_n-\frac{\hbar^2k_m^2}{2m^*})
\langle\zeta^m|\psi_{ni}\rangle\zeta^n ,
\end{equation}
where $i$ and $m$ ($i,m=0,1,$ and 2) indicate the initial and final
state of the system, respectively.

\begin{table*}
\caption{\footnotesize{Tangent of the phase shift of different
partial waves $l=0$, 1, 2, 3, and 4 for the first six iterations
within the MCF for an incident electron of kinetic energy $E_0$=0.6
meV. The number between bracket represents the power of ten,
\it{e.g.}, $(-4)=10^{-4}$.}}\label{tab1}
\begin{ruledtabular}
\begin{tabular}{ccccccccc}
 {iteration}& 0& 1& 2& 3& 4& 5& 6\\\hline
  $\tan\Delta_0$ &  -7.2103&  -6.0844 &  -1.7589 &  -1.6348
  &  -1.6442 &   -1.6347 &  -1.6347 \\
   $\tan\Delta_1 $&  -0.8833 &  -0.6862 &  -0.6874 &  -0.6305
  &  -0.6305 &  -0.6305&  -0.6305\\
  $\tan\Delta_2 $&  -0.0288
& 0.1838& 0.2465& 0.2495& 0.2495& 0.2495& 0.2495\\
  $ \tan\Delta_3 $&  5.43(-4) &  -2.62(-4)
&  -3.66(-5) & -3.64(-5) &  -3.64(-5)
 &  -3.64(-5) &  -3.64(-5) \\
   $\tan\Delta_4 $&  1.23(-4)&  -1.20(-4)&  -1.20(-4) &  -1.20(-4)
   &  -1.20(-4) &  -1.20(-4)
& -1.20(-4) \\
\end{tabular}
\end{ruledtabular}
\end{table*}

According to conservation of the total energy of the system, the
relation between the kinetic energies of the incident (scattered)
electron and the energies of the QD is given as
\begin{equation}\label{3ce}
\varepsilon_0+\frac{\hbar^2k_0^2}{2m^*}=\varepsilon_1+
\frac{\hbar^2k_1^2}{2m^*}= \varepsilon_2+\frac{\hbar^2k_2^2}{2m^*}.
\end{equation}
The corresponding L-S equation is reduced to
\begin{eqnarray}\label{lsoe}
&& \psi_{mi}(\mathbf{r})=\varphi_{i}(\mathbf{r})\delta_{mi}+\nonumber\\
&& +\sum_{n=0}^2\int d\mathbf{r}'
G_0(\mathbf{k}_m,\mathbf{r},\mathbf{r}')V_{mn}(\mathbf{r}')
\psi_{ni}(\mathbf{r}'),
\end{eqnarray}
where $ V_{mn}=V^{\rm st}_{mn}\pm V^{\rm ex}_{mn}$. The different
channels are coupled through the potential matrix elements $V_{mn}$
of the same total energy. In other words, the scattering for an
incident electron of momentum $k_0$ (associated to the QD ground
state $\varepsilon_0$) couples to that for an incident electron of
momentum $k_1$ (associated to the first excited QD state
$\varepsilon_1$) satisfying Eq.~(\ref{3ce}). Because the two
electrons can form both the singlet ($+$) and triplet ($-$) states,
the scattering cross-sections are different for these two distinct
cases:
\begin{equation}\label{sigos}
\sigma^{ {\rm s}}_{mi}(\theta )=\frac{k_m}{k_i^2}
|f^{(+)}_{k_m,k_i}(\theta)|^2
\end{equation}
for the singlet state, and
\begin{equation}\label{sigot}
\sigma^{ {\rm t}}_{mi}(\theta )=\frac{k_m}{k_i^2}
|f^{(-)}_{k_m,k_i}(\theta)|^2
\end{equation}
for the triplet states, where the scattering amplitudes are given by
\begin{eqnarray}\label{oeamp}
&&f^{(\pm)}_{k_m,k_i}(\theta)=-\frac{1}{4}\sqrt{\frac{2}{\pi}}
\sum_{n=0}^2\int d\mathbf{r}' e^{- i\mathbf{k}_m.\mathbf{r'}}
\nonumber \\
&&\times [V^{\rm st}_{mn}(\mathbf{r'})\pm V^{\rm
ex}_{mn}(\mathbf{r'})] \psi_{ni}(\mathbf{r'}).
\end{eqnarray}
The total differential cross-section or the spin-unpolarized (su)
DCS is determined by a statistical admixture of the singlet and
triplet state scattering,
\begin{equation}
\sigma^{ {\rm su}}_{mi}(\theta )=\frac{1}{4}\left(\sigma^{ {\rm
s}}_{mi}(\theta )+3\sigma^{ {\rm t}}_{mi}(\theta
)\right),\label{dcssu}
\end{equation}
where the factor 3 in the equation is due to statistical weight of triplet
states. Another interesting quantity is the spin-flip (sf) DCS which
describes the spin-flip scattering probability of an incident
electron resulting from the exchange interaction\cite{qct}. The sf-DCS
is given by,
\begin{equation}\label{dcssf}
\sigma^{\rm sf}_{mi}(\theta )=\frac{k_m}{4k_i^2}\left|f^{\rm
sf}_{k_m,k_i}(\theta )\right|^2 ,
\end{equation}
where
\begin{eqnarray}
&&f^{\rm sf}_{k_m,k_i}(\theta)=f^{(+)}_{k_m,k_i}(\theta
)-f^{(-)}_{k_m,k_i}(\theta )\nonumber \\
&&=-\frac{1}{2}\sqrt{\frac{2}{\pi}}\sum_{n=0}^2\int
d\mathbf{r}'e^{-i\mathbf{k}_m.\mathbf{r'}} V^{\rm
ex}_{mn}(\mathbf{r'}) \psi_{ni}(\mathbf{r'}).
\end{eqnarray}

\section{Numerical results and discussions}
We model the confinement potential of the QD by a 2D finite
parabolic potential
\begin{equation}\label{vqd}
V_{\rm QD}(r)=\left\{
\begin{array}{cc}
\frac{1}{2}m^*\omega_0^2(r^2-r_0^2)&,\;\; r<r_0 \\
0 &, \;\; r>r_0 ,
\end{array}
\right.
\end{equation}
where $\omega_0$ is the confinement frequency and $r_0$ is the
radius (size) of the dot. We will calculate in this section the
scattering due to a one-electron quantum dot. For such a system, the
solution of Eq.~(\ref{hqdN1}) is straightforward. We expand the
eigenfunction $\zeta^n$ in the Fock-Darwin basis\cite{fd} and
diagonalize numerically the Hamiltonian. The eigenstates can be
labeled by a set of quantum numbers $n$=($j,l$) with the radial
quantum number $j$=0,1,... and the angular momentum quantum number
$l$=0, $\pm$1,... . The state ($j=0,l=0$) is the ground state $n=0$
and the first two excited states ($j=0, l=\pm 1$) are degenerate
corresponding to $n=1$ and $n=2$ ($\varepsilon_1=\varepsilon_2$).

In order to solve the L-S equations, we use the partial wave
expansion in two dimensional system combined with the MCF. All the
involved functions are expanded in the angular momentum basis so
that we obtain a radial L-S equation for each angular momentum.
Numerically, we are able to choose the components of the angular
momentum which contribute to the cross-sections up to a desirable
precision. In Appendix B we show how the partial wave expansion can
be applied to the multichannel L-S equations in a two-dimensional
system.

\begin{figure*}[t!]
\begin{center}
\includegraphics[angle=0,scale=0.8]{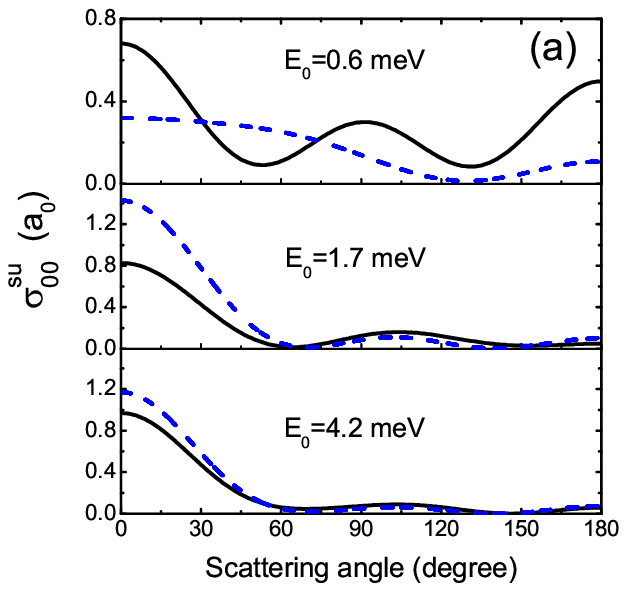}
\includegraphics[angle=0,scale=0.8]{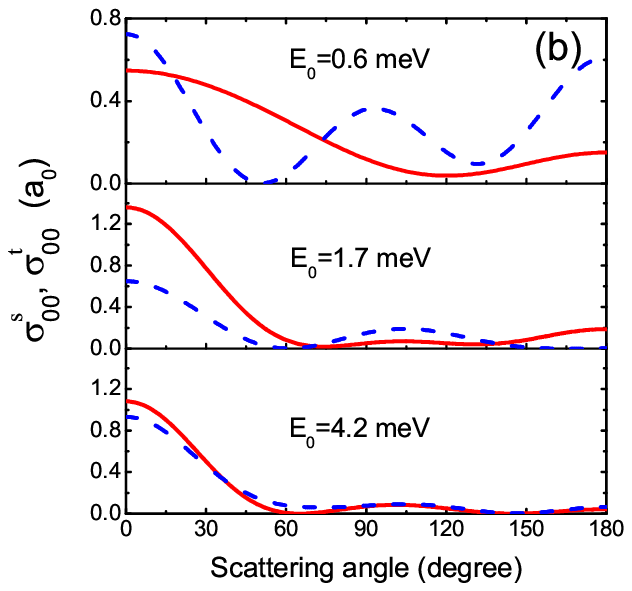}
\includegraphics[angle=0,scale=0.8]{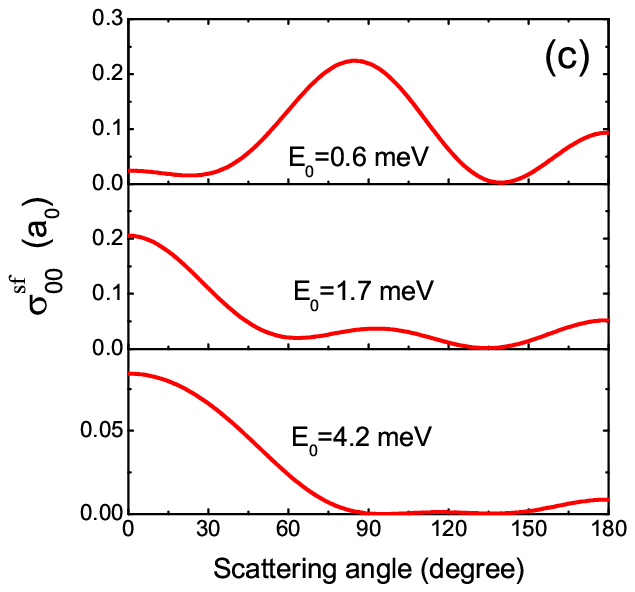}
\caption{\footnotesize{(Color online) The elastic DCS's obtained
within the one-channel model for electron scattering by the
one-electron QD of $\hbar\omega_0 =5$ meV and $r_0=35$ nm. The
incident electron energies are indicated in the figures. (a) The
spin-unpolarized DCS with (the solid curves) and without (the dashed
curves) the exchange potential; (b) The DCS due to the singlet state
(the solid curves) and the triplet state (the dashed curves); and
(c) The spin-flip DCS.}}\label{sce3}
\end{center}
\end{figure*}

\subsection{Convergency of the MCF}
The MCF was applied to the electron-atom scattering\cite{lee} and
electron-molecule scattering \cite{lee1}. In all those cases, it has
shown a rapid convergency. Here, we apply the MCF to the electron-QD
scattering in a two-dimensional semiconductor nanostructure. First
of all, we check the convergency of this method for electron
scattering through a QD. We consider a one-electron QD with
$\hbar\omega_0=5$ meV, $r_0= 35$ nm and an incident electron of
kinetic energy $E_0=\hbar^2k_0^2/2m^*=$ 0.6 meV. The results are
obtained within the one-channel approximation. For simplicity, only
the static scattering potential is considered and the exchange
potential is neglected in Eq.~(\ref{oeamp}). Table I gives the
calculated partial wave phase-shifts, for angular momenta up to $l =
4$, of the first six iterations. Because the localization length of
the confined electron wave-function in the QD is about
$a_0=\sqrt{\hbar/m^*\omega_0}$, the screening parameter is taken as
$\lambda= a_0^{-1}$ throughout this paper. ($a_0 = 14.75$ nm for a
GaAs QD of $\hbar\omega_0=5$ meV). We have performed calculations
with different values of $\lambda$. The calculated results have
showed that, although a smaller $\lambda$ has enhanced the static
potential scattering, the exchange effects and the spin-flip
scattering are not affected significantly. From Table I we can see
that the phase-shifts converge  at the fifth iteration. It also
shows that the first Born approximation, which corresponds to our
zero-{\it th} iteration calculation, is indeed a very poor
approximation in dealing with the electron-QD scattering. In order
to obtain a correct scattering cross-section through a QD, it is
necessary to use a robust method such as the MCF. Although the
results in Table I are for a particular case, we have verified that
all our calculations (with or without exchange interaction) in the
present paper are convergent within 6 iterations.

\begin{figure*}[t!]
\begin{center}
\includegraphics[angle=0,scale=1.0]{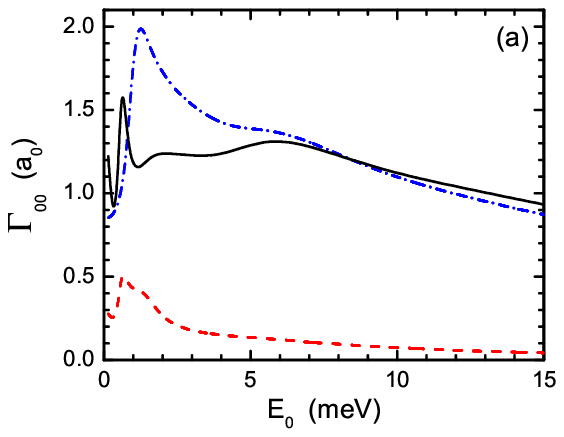}
\includegraphics[angle=0,scale=1.0]{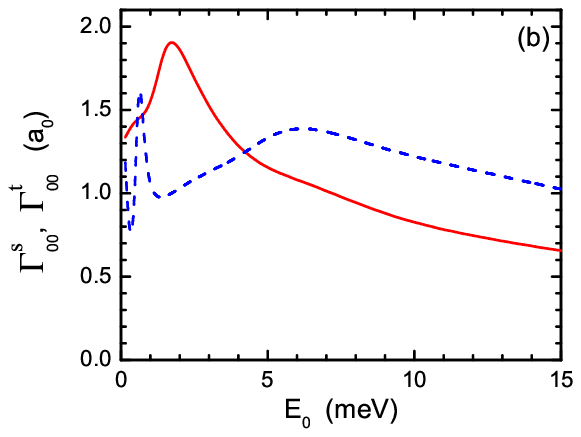}
\caption{\footnotesize{(Color online) The elastic ICS as a function
of $E_0$ for the one-electron QD. (a) The su-ICS (the solid curve),
the sf-ICS (the dashed curve), and the ICS due to the static
potential only (the dotted-dash curve); (b) The ICS due to the
singlet (the solid curve) and triplet states (the dashed curve)
during the scattering.}}\label{scte3}
\end{center}
\end{figure*}

\begin{figure*}[t!]
\begin{center}
\includegraphics[angle=0,scale=1.0]{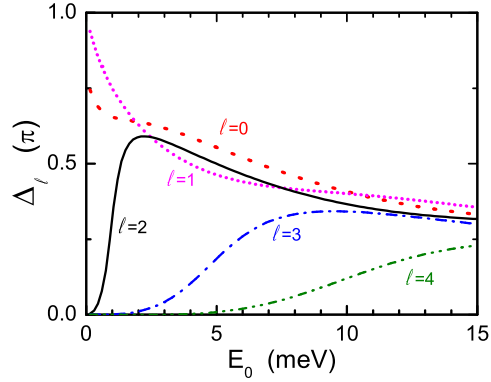}
\caption{\footnotesize{(Color online) The phase-shift as a function
of $E_0$ for the partial waves $l=0$, 1, 2, 3, and 4 due to the
static potential scattering of the one-electron QD.}}\label{scte3}
\end{center}
\end{figure*}

\subsection{One-channel scattering}
Within the one-channel approximation, we calculated the elastic DCS
for electron scattering by the one-electron QD of $\hbar \omega_{0}
=5$ meV and $r_0=$35 nm at incident kinetic energies $E_0=$0.6, 1.7,
and 4.2 meV. The obtained DCS's are presented in Fig. 1 as a
function of the scattering angle. Fig. 1(a) shows the total or the
su-DCS $\sigma^{\rm su}_{00}(\theta )$. To illustrate the effect of
exchange interaction, the su-DCS's due to the static potential only
({\it i.e.}, neglecting the exchange potential in Eqs.~(\ref{eses})
and (\ref{oeamp})) are given by the dashed curves in the figure. We
see that the exchange interaction is of significant contribution to
the low-energy and/or small angle scattering. The exchange effect on
the scattering is originated from the two different coupling states
between the incident and the QD electrons ({\it i.e.}, the singlet
and the triplet states) during the collision as indicated in
Eq.~(\ref{eses}). The corresponding DCS's due to the singlet
[$\sigma^{ {\rm s}}_{00}(\theta )$] and triplet states [$\sigma^{
{\rm t}}_{00}(\theta )$] defined by Eqs.~(\ref{sigos}) and
(\ref{sigot}), respectively, are shown in Fig.~1(b). In Fig. 1(c),
we plot the spin-flip (sf) DCS $\sigma^{ {\rm sf}}_{00}(\theta )$
given by Eq.~(\ref{dcssf}). Comparing Fig.~1(a) with Fig.~1(b), one
can see that the exchange interaction affects more strongly the
su-DCS when the difference between $\sigma^{ {\rm s}}_{00}(\theta )$
and $\sigma^{ {\rm t}}_{00}(\theta )$ is large. In Fig. 1(c) we
observe that the spin-flip scattering due to exchange potential
occurs mostly around $\theta \sim 90^o$ at lower incident energy
($E_0=$0.6 meV). For higher energies ($E_0=$1.7 and 4.2 meV), the
spin-flip scattering is more relevant at small scattering angles.

Fig.~2(a) shows the ICS as a function of incident electron energy
$E_{0}$ for the spin-unpolarized scattering and for that considering
the static potential only. We see that the exchange interaction
affects significantly the ICS at low $E_{0}$. At higher energies,
however, the ICS is dominated by the static potential. In Fig.~2(a)
we also present the spin-flip ICS (the dashed curve). A maximum
spin-flip probability is found at $E_0=$1.1 meV which is about
$37\%$ of the total scattering. In Fig.~2(b), we plot the ICS due to
the singlet and the triplet states. It shows a strong dependence of
the ICS on the spin states of two electrons in the system.

The scattering peaks in ICS due to the static potential (the
dotted-dash curve in Fig.~2(a)) at $E_{0}=1.22$ and 6.0 meV are due to
 the occurrence of the
so-called shape resonances, resulting from a virtual confined state
at the corresponding energy. In order to clarify the origin of these
features, we plot in Fig.~3 the corresponding partial wave
phase-shifts $\Delta_l$ (for $l=0$, 1, 2, 3 and 4) due to the static
potential. At $E_0\to 0$, $\Delta_0$ and $\Delta_1$ are larger than
$\pi/2$ indicating the presence of the bound states of angular
momenta $l=$0 and 1 in the QD. A rapid increase of $\Delta_2$
($\Delta_3$) at around $E_0=1.2$ meV (6.0 meV) corresponds to a
virtual bound state of $l=2$ ($l=3$) leading to the shape resonance
scattering peak in the ICS. Similarly, peaks in the ICS at
$E_{0}=1.72$ meV (0.57 meV) for the singlet (triplet) state
scattering in Fig.~2(b) can be related to the virtual bound states
in the system. The broad peak in the ICS of triplet state is
possibly result from a virtual state of two interacting electrons.

\begin{figure*}[t!]
\begin{center}
\includegraphics[angle=0,scale=1.2]{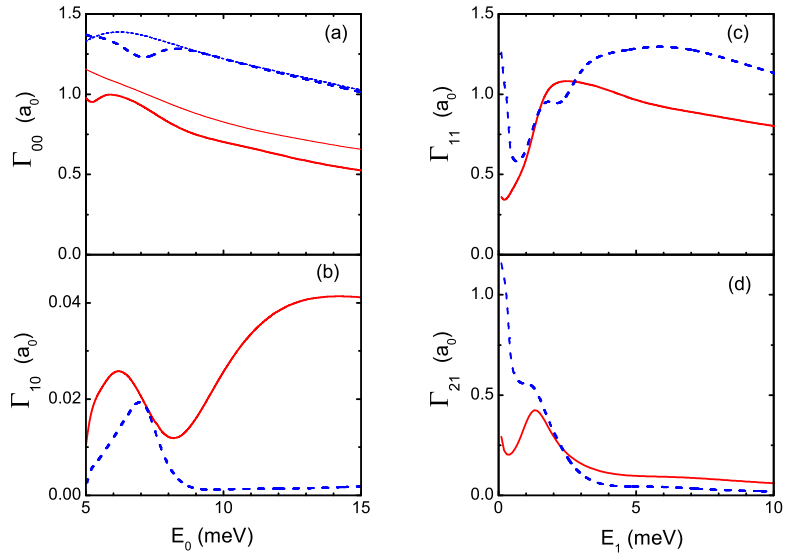}
\caption{\footnotesize{(Color online) The multichannel integral
cross-sections $\Gamma^{\rm s}_{mi}$ (the solid curves) and
$\Gamma^{\rm t}_{mi}$ (the dashed curves) as a function of $E_i$
($i=0,1$) for the one-electron QD considering the three lowest
energy states. The thin curves in (a) are the corresponding results
within the one-channel model.}}
\end{center}
\end{figure*}

\subsection{Multi-channel scattering}
The energy difference between the first excited state and the ground
state is 4.90 meV in the QD of $\hbar\omega_0$=5 meV and $r_0$=35
nm. When the kinetic energy $E_0$ of an incident electron is higher
than this energy difference, the inelastic scattering channel is
open which leaves the QD in the excited state $\varepsilon_1$ after
scattering. In such a case, the multichannel scattering process has
to be considered. When the three lowest states of a one-electron QD
are included in the calculation, there are 9 possible scattering
channels. For the present QD, as the first excited state is two-fold
degenerate, {\it i.e.} $\varepsilon_1=\varepsilon_2$, we find the
following scattering cross-sections: the elastic $\Gamma_{00}(E_0)$
and inelastic scattering $\Gamma_{10}(E_0)$ for the QD initially in
its ground state; the elastic scattering $\Gamma_{11}(E_1)$ and
super-elastic scattering $\Gamma_{01}(E_1)$ for the QD initially in
the first excited state. There are also two excitation
cross-sections $\Gamma_{21}(E_1)=\Gamma_{12}(E_1)$ for the QD in the
excited states of different angular momenta, although in these cases
the energy of the scattered electron is the same as that of the
elastic scattering $\Gamma_{11}(E_1)$, {\it i.e.}
$\varepsilon_1=\varepsilon_2$ and $k_1=k_2$ in Eq~(\ref{3ce}). In
Fig.~(4) we show the different integral cross-sections due to the
singlet and the triplet states. Figs.~4(a) and 4(b) give the elastic
ICS ($\Gamma_{00}$) and inelastic ICS ($\Gamma_{10}$), respectively,
for the QD initially in its ground state. The inelastic
cross-section is two orders of magnitude smaller than the elastic
one. Moreover, the inelastic scattering due to the triplet state is
very weak at higher energies. Coupling between different QD levels
(or different scattering channels) leads to resonant scattering on
both the elastic and inelastic cross-sections. The thin curves in
Fig.~4(a) are the previous results in Fig.~2(b) of the ICS within
the one-channel model. We see that the one-channel approximation
yields quite good results for the elastic scattering.

Figs.~4(c) and 4(d) show the elastic ICS for the QD in the first
excited state. At small incident energy $E_1$, the scattering due to
the triplet state is much stronger than that due to the singlet
state in this case. If the QD remains in the same excited state
after the scattering, the ICS $\Gamma_{11}$ (=$\Gamma_{22}$) as
shown in Fig.~4(c) is large at higher energy and decreases slowly
with increasing $E_1$. However, if the angular momentum changes
after the scattering, the ICS $\Gamma_{21}$ (=$\Gamma_{12}$)
vanishes rapidly (both due to the singlet and triplet states) at
high incident energies.

\section{Conclusion remarks}
We presented a general method to calculate the electron scattering
through an N-electron QD embedded in a two-dimensional semiconductor
system. The multichannel L-S equations are solved numerically using
the iterative method of continued fractions considering the
electron-electron interactions. We applied this method to the case
where only one electron is confined in the QD. The results indicate
a rapid convergency of this method for two-dimensional scattering in
a semiconductor nanostructure. It also shows that the first Born
approximation is so poor that cannot yield correct scattering
cross-section.

We found that the exchange effects are relevant when the kinetic
energy of incident electron is small, as showed by the obtained DCS
and ICS. The shape resonances were found in the elastic ICS
including or not the exchange potential. The spin-flip cross-section
due to exchange interaction shows a maximum both in the DCS as a
function of the scattering angle and in the ICS as a function of the
incident electron energy. The maximum spin-flip scattering reaches
as high as more than 30\% in comparison to the total scattering. In
multichannel scattering including the excited states of the QD, we
obtained the inelastic scattering cross-sections. They are about two
orders of magnitude smaller than the elastic ones.

In this paper, we emphasize the theoretical approach and numerical
method to calculate the electron scattering by a charged quantum
dot. The scattering cross-sections were obtained for a spin
unpolarized system. It can be extended to a spin polarized system
which is of great interest for electron transport in semiconductor
nanostructures. The application to a spin polarized system is
straightforward as soon as the initial spin states of the system are
defined. In the numerical calculation, we presented the
cross-sections due to a one-electron QD scattering. For a QD of two
or more electrons, we need firstly to know the eigenstates of the QD
with electron-electron interactions. Then the scattering
cross-sections can be calculated according to the total wavefunction
defined by Eq.~(\ref{psi}) as what has been done for electron-atom
and electron-molecule scattering where several electrons are
presented.\cite{lee1}

\acknowledgments  This work was supported by FAPESP and CNPq
(Brazil). One of the authors (L. K. C.) would like to thank K. T.
Mazon for stimulating discussions.

\appendix
\section{Method of Continued Fractions}\label{mcf}

The method of continued fractions (MCF)\cite{mcf} is an iterative
method to solve the L-S equation. To apply this method for a
multi-channel scattering we have firstly to rewrite
Eq.~(\ref{elsmc}) in a matrix form:
\begin{equation}\label{elsmcm}
\widetilde{\Psi}=\widetilde{\varphi}+\widetilde{G}^{(0)}\widetilde{V}
\widetilde{\Psi}.
\end{equation}
In the first step to start the MCF, we use the scattering potential
$\widetilde{V}=V^{(0)}$ and the free electron wave-function
$\widetilde{\varphi}=|\varphi^{(0)}\rangle$ in Eq.~(\ref{elsmcm}).
Afterwards, we define the n$th$-order weakened potential as
\begin{equation}\label{un}
V^{(n)}=V^{(n-1)}- \frac{V^{(n-1)}|\varphi^{(n-1)}\rangle\langle
\varphi^{(n-1)}|V^{(n-1)}}{\langle
\varphi^{(n-1)}|V^{(n-1)}|\varphi^{(n-1)}\rangle},
\end{equation}
where
\begin{equation}
|\varphi^{(n)}\rangle=\widetilde{G}^{(0)}V^{(n-1)}|\varphi^{(n-1)}\rangle.
\end{equation}
The n$th$-order correction of the \textbf{T} matrix can be obtained
through
\begin{eqnarray}\label{tma}
T^{(n)}=\langle \varphi^{(n-1)}|V^{(n-1)}|\varphi^{(n)}\rangle+
\langle \varphi^{(n)}|V^{(n)}|\varphi^{(n)}\rangle && \nonumber\\
\times\left[\langle \varphi^{(n)}|V^{(n)}|\varphi^{(n)}\rangle-
T^{(n+1)}\right]^{-1}\langle
\varphi^{(n)}|V^{(n)}|\varphi^{(n)}\rangle. &&
\end{eqnarray}
Hence, we can stop the iteration when the potential $V^{(N)}$
becomes weaker enough. In the numerical calculation, we start with
$T^{(N+1)}=0$ and evaluate $T^{(N)},T^{(N-1)},...,$ and $T^{(1)}$.
Therefore the \textbf{T} matrix is given by
\begin{equation}\label{tma}
\mathbf{T}=\langle \varphi^{(0)}|V^{(0)}|\varphi^{(0)}\rangle+
T^{(1)}\frac{\langle \varphi^{(0)}|V^{(0)}|\varphi^{(0)}\rangle}{\langle
\varphi^{(0)}|V^{(0)}|\varphi^{(0)}\rangle-T^{(1)}} .
\end{equation}

\section{Partial wave expansion}\label{secop}

In two dimensions the angular momentum basis is given by\cite{2d},
\begin{equation}\label{ma}
\Theta_l(\phi)=\sqrt{\frac{\kappa_l}{2\pi}}\cos(l\phi)
\end{equation}
where $l=0,1,2,...$, $\kappa_l=2$ for $l\neq0$ and $\kappa_l=1$ for
$l=0$. In applying the partial wave expansion in the multi-channel
scattering problem Eq.~(\ref{elsmc}), we expand all functions, {\it
i.e.}, the incident free electron wavefunction
$\varphi_{i}(\mathbf{r})$, the Green's function
$G^{(0)}(\mathbf{k}_m,\mathbf{r},\mathbf{r'})$, and the scattered
electron wavefunction $\psi_{mi}(\mathbf{r})$, in the angular
momentum basis as follows,
\begin{equation}
\varphi_{i}(\mathbf{r})=\sum_{l,l'=0}^{\infty}
\sqrt{\frac{\kappa_l}{2\pi}}i^lJ_l(kr)\delta_{ll'}
  \Theta_l(\phi_r)\Theta_{l'}(\phi_k),
\end{equation}
and
\begin{equation}
\psi_{mi}(\mathbf{r})=\sum_{l,l'=0}^{\infty}\psi_{mi}^{l,l'}(k,r)
\Theta_l(\phi_r)\Theta_{l'}(\phi_k),
\end{equation}
where $\phi_r$ and $\phi_k$ are the variables due to expansion on
the position $\bf r$ and momentum ${\bf k}$, respectively. The
expansion on the Green's function yields the following expression,
\begin{eqnarray}
&& G^{(0)}(\mathbf{k}_m,\mathbf{r},\mathbf{r'}) =
\\ && -\frac{i\pi}{2}\sum_{l=0}^{\infty}\sqrt{\frac{\kappa_l}{2\pi}}J_l(k_mr_<)
H^{(1)}_l(k_mr_>) \Theta_l(\phi_r)\Theta_l(\phi_{r'}), \nonumber
\end{eqnarray}
where $k=k_i$, $r_<={\rm min}(r,r')$, $r_>={\rm max}(r,r')$,
$J_l(k_mr)$ ($Y_l(k_mr$)) is the Bessel (Neumann) function and
$H^{(1)}_l(k_mr)=J_l(k_mr)+iY_l(k_mr)$ is the Hankel function
\cite{morse}. Using the partial wave expansion the
Lippmann-Schwinger equation can be reduced to a set of radial
equations. The radial Lippmann-Schwinger equation corresponding to
Eq.~(\ref{elsmc}) is given by,
\begin{eqnarray}
&&\psi_{mi}^{l,l'}(k,r)=\sqrt{\frac{\kappa_l}{2\pi}}i^lJ_l(kr)
\delta_{ll'}\delta_{mi}\label{elsr}\\
&&+\sum_{l''=0}^\infty\sum_{n=0}^\infty\int_0^\infty
r'dr'g^l_0(k_m,r,r')V_{mn}^{l,l''}(r')\psi_{ni}^{l'',l'}(r'),\nonumber
\end{eqnarray}
where
\begin{equation}\label{greenm}
g^l_0(k_m,r,r')=\frac{-i\pi}{2}\sqrt{\frac{\kappa_l}{2\pi}}J_l(k_mr_<)
H^{(1)}_l(k_mr_>)
\end{equation}
and
\begin{equation}\label{vpar}
V^{l,l''}_{mn}(r')=\int_0^{2\pi}d\phi_{r'}
\Theta_l(\phi_{r'})V_{mn}(\mathbf{r'})\Theta_{l''}(\phi_{r'}).
\end{equation}
We see that, when the partial wave method is used, there is a change
in the continuum variable $\phi$ to a partial wave $l$.
Consequently, the wavefunction $\psi_{mi}(\mathbf{r})$ becomes a
matrix function with elements $\psi^{l,l'}_{mi}(k,r)$.

The partial wave expansion for the exchange potential is a little
subtle due to its non-locality. Here, we show some details how the
partial wave expansion is applied in this case. We take as an
example the exchange potential which couples the channels $n$ and
$m$ for a single electron spin-orbital $\alpha$,
\begin{equation}
V_{mn}^{\rm ex}(\mathbf{r})\psi_{ni}(\mathbf{r})
=-\frac{e^2}{\epsilon^*_0} \zeta^n_\alpha(\mathbf{r})\int
d\mathbf{r_1} \zeta_\alpha^{m*}(\mathbf{r_1})\frac{1}
{|\mathbf{r}-\mathbf{r_1}|}\psi_{ni}(\mathbf{r_1}) . \label{vex2d}
\end{equation}
The partial wave expansion of the spin-orbital function is given by
\begin{equation}
\zeta^n_{\alpha}(\mathbf{r})=
\sum_{l=0}^\infty\zeta^{l}_{n\alpha}(r)\Theta_l(\phi_r).
\end{equation}
The product of two different functions can also be expanded in the
angular momentum basis as follows,
\begin{equation}
\psi_{ni}(\mathbf{r})\zeta^{m*}_{\alpha}(\mathbf{r})=\sum_{l,l'}
\Pi_{ni;m\alpha}^{l,l'}(r)\Theta_l(\phi_r)\Theta_{l'}(\phi_k),
\end{equation}
where
\begin{eqnarray}
&&\Pi_{ni;m\alpha}^{l,l'}(r) \\
&&
=\sum_{\lambda,\lambda'}\frac{\psi_{ni}^{\lambda,l'}(k,r)
\zeta^{\lambda'*}_{m\alpha}(r)}{2\sqrt{2\pi}}
\sqrt{\frac{\kappa_\lambda\kappa_{\lambda'}}{\kappa_l}}
\left(\delta_{l,\lambda+\lambda'}
+\delta_{l,|\lambda-\lambda'|}\nonumber \right).
\end{eqnarray}
Using the above relation, we obtain Eq.~(\ref{vex2d}) in the partial
wave expansion form,
\begin{eqnarray}\label{vex2d3}
&&V_{mn}^{\rm
ex}(\mathbf{r})\psi_{ni}(\mathbf{r})=-\frac{e^2}{\epsilon^*_0}
\zeta^n_\alpha(\mathbf{r})\sum_{l,l'}\Theta_l(\phi_{r})\Theta_{l'}(\phi_k)
\\
&&\times \int_0^\infty r_1dr_1\Pi_{ni;m\alpha}^{l,l'}(r_1)
\int_{0}^{2\pi} \frac{\Theta_l(\theta)d\theta}
    {\sqrt{r^2+r_1^2-2rr_1\cos(\theta)}} , \nonumber
\end{eqnarray}
where $\theta=\phi_{r}-\phi_{r_1}$. To solve the angular integral we
use the generating function of the Legendre Polynomials\cite{morse},
\begin{equation}
\frac{1}{\sqrt{r^2+r_1^2-2rr_1\cos(\theta)}}=
\sum_{j=0}^\infty\frac{r_<^j}{r_>^{j+1}}P_j(\cos\theta)\label{gepl},
\end{equation}
where $r_<={\rm min}(r,r_1)$, $r_>={\rm max} (r,r_1)$ and
$P_j(\cos\theta)$ are the Legendre Polynomials. Thus the angular
integral that we need to solve is
\begin{eqnarray}
c_{l,j}=\int_{0}^{2\pi}d\theta \Theta_l(\theta)P_j(\cos\theta).
\label{vexang}
\end{eqnarray}
Substituting the Eqs.~(\ref{gepl}) and (\ref{vexang}) into
Eq.~(\ref{vex2d3}) we obtain finally the exchange potential
\begin{eqnarray}
&&V_{mn}^{\rm
ex}(\mathbf{r})\psi_{ni}(\mathbf{r})=-\frac{e^2}{\epsilon^*_0}
\zeta^n_\alpha(\mathbf{r})\sum_{l,l'}\Theta_l(\phi_{r})
\Theta_{l'}(\phi_k) \times\nonumber\\&&\times \sum_{j=0}^\infty
\int_0^\infty r_1dr_1\Pi_{ni;m\alpha}^{l,l'}(r_1)
 c_{l,j}\frac{r_<^j}{r_>^{j+1}}. \label{vex2d4}
\end{eqnarray}
In the numerical calculations, we firstly evaluate the coefficients
$c_{l,j}$ given by Eq.~(\ref{vexang}). Then the integration on $r_1$
in Eq.~(\ref{vex2d4}) is performed for each iteration in the MCF.
Finally we multiply the result by
$-\frac{e^2}{\epsilon^*_0}\zeta^n_\alpha(\mathbf{r})$.

Within the one-channel approximation ($i=m=n=0$), the calculations
can be further simplified by using the concept of phase shift.
Considering a central potential $V(r)$ ($l=l'=l''$),
Eq.~(\ref{elsr}) becomes
\begin{equation}
\psi^{l}(k,r)=\sqrt{\frac{\kappa_l}{2\pi}}i^lJ_l(kr)+\int_0^\infty
r'dr'g^l_0(k,r,r')V(r')\psi^{l}(k,r')\label{elsvr}
\end{equation}
where $\psi^{l}(k,r)=\psi^{l,l}_{00}(k,r)$. To define the
phase-shift we write the asymptotic form of the above equation as
\begin{equation}\label{elsvrdl}
\psi^{l}(k,r)\hbox{\space \raise-2mm\hbox{$\textstyle
\longrightarrow \atop \scriptstyle r\rightarrow\infty$}
\space}A_l\sqrt{\frac{1}{
kr}}\cos(kr-\frac{l\pi}{2}-\frac{\pi}{4}-\Delta_l),
\end{equation}
where $\Delta_l$ is the phase-shift. Comparing the coefficients of
$e^{ikr}$ and $e^{-ikr}$ of Eq.~(\ref{elsvrdl}) with the asymptotic
form of Eq.~(\ref{elsvr}) one can obtain the following relations
\begin{equation}
A_l=2\sqrt{\frac{\kappa_l}{\pi}}i^le^{i\Delta_l},
\end{equation}
and
\begin{equation}
e^{i\Delta_l}\sin\Delta_l=\frac{-\pi}{2i^l}\int_0^\infty
r'dr'J_l(kr')V(r')\psi^{l}(r').
\end{equation}
On the other hand, from the definition of the scattering amplitude
in Eq.~(\ref{elsmcass}), we can express the scattering amplitude
$f_{k_0,k_0}$ in terms of the phase-shift\cite{2d} $\Delta_l$,
\begin{equation}
f_{k_0,k_0}(\theta)=2\sum_{l=0}^\infty
\sqrt{\frac{\kappa_l}{\pi}}e^{i\Delta_l}\sin\Delta_l
\Theta_l(\theta).
\end{equation}
The corresponding DCS is $\sigma_{00}
(\theta)=\left|f_{k_0,k_0}(\theta)\right|^2/k $ and the ICS is given
by
\begin{equation}
\Gamma_{00}=\frac{4}{k}\sum_{l=0}^\infty
\kappa_l\sin^2\Delta_l\label{scps}.
\end{equation}

\end{document}